\documentclass[aps,twocolumn,superscriptaddress,showpacs]{revtex4-1}

\usepackage{graphicx}
\usepackage{dcolumn}
\usepackage{bm}
\usepackage{amsmath}
\usepackage{color}
\usepackage{ulem}
\usepackage{threeparttable}
\usepackage{enumerate}
\usepackage{paralist}

\begin{document}

\title{Thermodynamics and heat transport of quantum spin liquid candidates NaYbS$_2$ and NaYbSe$_2$}

\author{N. Li}
\affiliation{Anhui Provincial Key Laboratory of Magnetic Functional Materials and Devices, Institutes of Physical Science and Information Technology, Anhui University, Hefei, Anhui 230601, People's Republic of China}

\author{M. T. Xie}
\affiliation{Beijing National Laboratory for Condensed Matter Physics, Institute of Physics, Chinese Academy of Sciences, Beijing 100190, People's Republic of China}
\affiliation{School of Physical Science and Technology, Lanzhou University, Lanzhou 730000, People's Republic of China}

\author{Q. Huang}
\affiliation{Department of Physics and Astronomy, University of Tennessee, Knoxville, Tennessee 37996, USA}

\author{Z. W. Zhuo}
\affiliation{Beijing National Laboratory for Condensed Matter Physics, Institute of Physics, Chinese Academy of Sciences, Beijing 100190, People's Republic of China}
\affiliation{School of Physical Science and Technology, Lanzhou University, Lanzhou 730000, People's Republic of China}

\author{Z. Zhang}
\affiliation{Beijing National Laboratory for Condensed Matter Physics, Institute of Physics, Chinese Academy of Sciences, Beijing 100190, People's Republic of China}

\author{E. S. Choi}
\affiliation{National High Magnetic Field Laboratory, Florida State University, Tallahassee, FL 32310-3706, USA}

\author{Y. Y. Wang}
\affiliation{Anhui Provincial Key Laboratory of Magnetic Functional Materials and Devices, Institutes of Physical Science and Information Technology, Anhui University, Hefei, Anhui 230601, People's Republic of China}

\author{H. Liang}
\affiliation{Anhui Provincial Key Laboratory of Magnetic Functional Materials and Devices, Institutes of Physical Science and Information Technology, Anhui University, Hefei, Anhui 230601, People's Republic of China}

\author{Y. Sun}
\affiliation{Anhui Provincial Key Laboratory of Magnetic Functional Materials and Devices, Institutes of Physical Science and Information Technology, Anhui University, Hefei, Anhui 230601, People's Republic of China}

\author{D. D. Wu}
\affiliation{Anhui Provincial Key Laboratory of Magnetic Functional Materials and Devices, Institutes of Physical Science and Information Technology, Anhui University, Hefei, Anhui 230601, People's Republic of China}

\author{Q. J. Li}
\affiliation{School of Physics and Optoelectronics, Anhui University, Hefei, Anhui 230061, People's Republic of China}

\author{H. D. Zhou}
\affiliation{Department of Physics and Astronomy, University of Tennessee, Knoxville, Tennessee 37996, USA}

\author{G. Chen}
\email{chenxray@pku.edu.cn}
\affiliation{International Center for Quantum Materials, School of Physics, Peking University, Beijing 100871, China}

\author{X. Zhao}
\email{xiazhao@ustc.edu.cn}
\affiliation{School of Physics Sciences, University of Science and Technology of China, Hefei, Anhui 230026, People's Republic of China}

\author{Q. M. Zhang}
\email{qmzhang@ruc.edu.cn}
\affiliation{Beijing National Laboratory for Condensed Matter Physics, Institute of Physics, Chinese Academy of Sciences, Beijing 100190, People's Republic of China}
\affiliation{School of Physical Science and Technology, Lanzhou University, Lanzhou 730000, People's Republic of China}

\author{X. F. Sun}
\email{xfsun@ahu.edu.cn}
\affiliation{Anhui Provincial Key Laboratory of Magnetic Functional Materials and Devices, Institutes of Physical Science and Information Technology, Anhui University, Hefei, Anhui 230601, People's Republic of China}
\affiliation{School of Physics Sciences, University of Science and Technology of China, Hefei, Anhui 230026, People's Republic of China}
\affiliation{Collaborative Innovation Center of Advanced Microstructures, Nanjing University, Nanjing, Jiangsu 210093, People's Republic of China}

\date{\today}

\begin{abstract}

We study the ultralow-temperature thermodynamics and thermal conductivity ($\kappa$) of the single-crystal rare-earth chalcogenides NaYbS$_2$ and NaYbSe$_2$, which have an ideal triangular lattice of the Yb$^{3+}$ ions and have been proposed to be quantum spin liquid candidates. The magnetic specific heat divided by temperature $C_{\rm{mag}}/T$ is nearly constant at $T <$ 200 mK, which is indeed the indication of the gapless magnetic excitations with a constant density of states. However, we observe a vanishingly small residual term $\kappa_0/T$, which points to the absence of mobile fermionic excitations in these materials. Both the weak temperature dependence of $\kappa$ and the strong magnetic-field dependence of $\kappa$ suggest the significant scattering between the spinons and phonons, which actually supports the existence of gapless or tiny-gapped quantum spin liquid. Moreover, the $\kappa(B)/\kappa(0)$ isotherms show a series of field-induced magnetic transitions for $B \parallel a$, confirming the easy-plane anisotropy, which is consistent with the results of ac magnetic susceptibility. We expect our results to inspire further interests in the understanding of the spinon-phonon coupling in the spin liquid systems.

\end{abstract}


\maketitle

\section{Introduction}

In low-dimensional quantum magnets, the strong frustration effect originated from the competing magnetic exchange interactions suppresses conventional long-range order, together with the small spin quantum number, leads to an exotic quantum disorder state without breaking any symmetry even at zero Kelvin -- quantum spin liquid (QSL). It was originally proposed by Anderson in 1973 based on two-dimensional triangular lattice antiferromagnet, and then applied to explain the mechanism of high-temperature superconductivity \cite{PWAnderson, Science321, ContempPhys, Nature464, Science367}. Since then, many theoretical and experimental efforts have been devoted to QSL. Theoretical studies have proposed several kinds of QSLs, including a gapless $U(1)$ Dirac QSL, a gapped $Z_2$ QSL, and chiral spin liquid etc \cite{RevModPhys89, RepPeogPhys80}. Experimentally, various types of QSL candidates have been discovered in two-dimensional triangular lattice, kagome lattice, honeycomb lattice or three-dimensional pyrochlore structures.

It is well known that spin-1/2 triangular-lattice antiferromagnets are ideal QSL candidates. EtMe$_3$Sb[Pd(dmit)$_2$]$_2$ and $\kappa$-(BEDT-TTF)$_2$Cu$_2$(CN)$_3$ are early found organic QSL candidate materials and have been widely studied \cite{Science328, PhysRevLett107001}. And then, a variety of inorganic materials have emerged, including Ba$_3$CuSb$_2$O$_9$, YbMgGaO$_4$, TbInO$_3$, NaBaCo(PO$_4$)$_2$, Pr$M$Al$_{11}$O$_{19}$ ($M =$ Mg, Zn), NdTa$_7$O$_{19}$, $AReCh_2$ ($A =$ alkali, $Re =$ rare earth, $Ch =$ O, S, Se) \cite{PhysRevLett147204, SciRep5, NatPhys262, NC4216, JAlloysCompd, JMaterChem, NatMater416, CPL35, PRB100-220407}. Among them, rare-earth-based spin frustrated materials have attracted much attention, in which the strong spin-orbit coupling of rare-earth ions can realize an effective $J_{eff} =$ 1/2 moment and strong quantum fluctuations stabilize QSL state. In particular, a 4$f$-based compound YbMgGaO$_4$ has been investigated extensively. The absence of long-range magnetic ordering down to several tens of milliKelvin, a $T^{2/3}$ behavior of magnetic specific heat \cite{SciRep5}, a small residual thermal conductivity $\kappa_0/T$ ($\sim$ 0.0058 W/K$^2$m) \cite{NC4949}, and a continuous magnetic excitation spectrum \cite{Nature540, NatPhys117} are attributed to the gapless QSL ground state with spinon Fermi surface. On the contrary, the frequency dependence of the ac susceptibility and a persistent excitation continuum at polarized-field \cite{NC4949, NatPhys117} question the QSL ground state. The random distribution of Mg$^{2+}$ and Ga$^{3+}$ ions may play a crucial role in the controversy on the magnetic ground state of YbMgGaO$_4$ \cite{PhysRevLett157201}. To better characterize the actual physics of QSL, it is necessary to discover new QSL candidates to eliminate disorder effects.

The Yb delafossites NaYb$Ch_2$ ($Ch =$ O, S, Se) were found to be the promising QSL materials, in which Yb$^{3+}$ ions form an ideal triangular lattice and free from mixing disorder. The magnetic properties of  NaYb$Ch_2$ can be systematically tuned by changing the chalogen ions (intralayer interactions). Previous studies reported the power-law temperature dependence of the magnetic specific heat and the continuous magnetic excitations observed by inelastic neutron scattering (INS), indicative of the gapless $U(1)$ QSL with spinon Fermi surface \cite{PhysRevB144432, QuantumFrontiers13, PRX021044}. The fingerprints of the gapless spinon Fermi surface are: (i) a broad and continuous magnetic excitations spectrum in INS experiment \cite{Nature492, Nature540, NatPhys117}; (ii) a power law temperature dependence of magnetic specific heat ($C \sim T^\alpha$) \cite{SciRep5, NatPhys459, NatCommun275}; (iii) a finite residual thermal conductivity $\kappa_0/T$ at zero-Kelvin \cite{Science328, NC4216, PhysRevB235124, PhysRevX031064, PhysRevResearch2}. Obviously, NaYb$Ch_2$ exhibit the first two hallmarks and lack the experimental studies of low-temperature thermal conductivity due to the challenge in growing high-quality single crystals. For NaYbO$_2$, all experimental researches are based on the polycrystalline samples and limit the further characterization. For NaYbS$_2$ and NaYbSe$_2$, the researchers have successfully grown single crystal samples, but they are so thin and fragile that the high-quality heat transport measurement is still very challenging. In this regard, one earlier work reported the ultralow-temperature $\kappa$ results of NaYbSe$_2$ single crystal \cite{TheInnovation}, which however displayed a puzzling peak at 5 T on the $\kappa(B)$ curves ($B \parallel c$) since all the existing experimental studies do not suggest any magnetic transition at this critical field.

In this work, we have grown high-quality NaYbS$_2$ and NaYbSe$_2$ single crystals and performed ultralow temperature thermal conductivity measurement, as well as the ultralow temperature ac magnetic susceptibility and specific heat measurements. In addition to confirming the absence of long-range magnetic order down to several tens of milliKelvin and the linear temperature dependent magnetic specific heat, we observed a vanishingly small residual term $\kappa_0/T$ for both materials, which indicates that there are no mobile fermionic excitations. Moreover, we observed a series of field-induced magnetic transitions for $B \parallel a$, including the 1/3 magnetization plateau associated with an up-up-down spin structure that demonstrates the easy-plane properties of NaYbS$_2$ and NaYbSe$_2$.

\section{Experiments}

High-quality NaYbS$_2$ and NaYbSe$_2$ single crystals were grown by the flux method \cite{CPL35, PhysRevB184419, PRB035144, SciChinaPhysMechAstron67}. Large and shinny thin-pallet-like crystals were obtained. The important progress of our crystal growth is that the thickness of single crystals reaches several tens micrometers, which guarantees the reliable heat transport measurements. The ac susceptibility was measured using the conventional mutual inductance technique (with a combination of ac current source and a lockin amplifier) at SCM1 dilution fridge magnet of the National High Magnetic Field Laboratory, Tallahassee \cite{PRB064401}. The typical ac field strength is 1.1--1.6 Oe. The specific heat was measured by relaxation technique using the Physical Property Measurement System (PPMS) (DynaCool, Quantum Design) with a dilution insert. The heat transport measurements were performed by using a ``one heater, two thermometers" method in a $^3$He/$^4$He dilution refrigerator (70 mK $< T <$ 1 K) and $^3$He refrigerator (0.3 K $< T <$ 30 K), equipped with a 14 T superconducting magnet. The samples were cut into dimensions of 2.13 $\times$ 1.56 $\times$ 0.052 mm$^3$ and 3.00 $\times$ 1.07 $\times$ 0.029 mm$^3$ for NaYbS$_2$ and NaYbSe$_2$, respectively, with the $c$ axis along the thickness direction. The heat currents ($J$) were along the longest dimensions in the $a$ axis, and the magnetic fields were applied along either the $c$ axis ($B \perp J$) or the $a$ axis ($B \parallel J$).

\section{Results and Discussion}

\subsection{Crystal structure and magnetic susceptibility}

\begin{figure}
\includegraphics[clip,width=8.5cm]{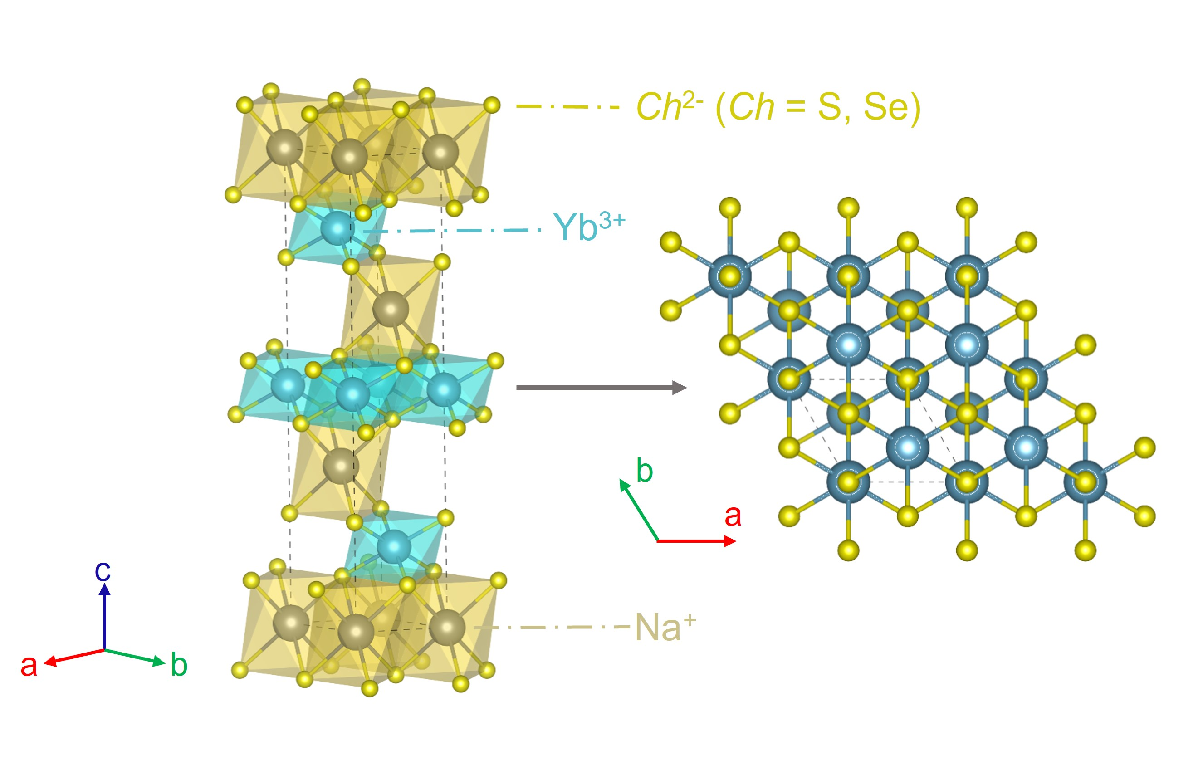}
\caption{Crystal structure of NaYbS$_2$ and NaYbSe$_2$ and the schematic illustration of the perfect triangular layer with the Yb$^{3+}$ ions.}
\label{S}
\end{figure}

Figure 1 shows the crystal structure of NaYbS$_2$ and NaYbSe$_2$, in which the magnetic Yb$^{3+}$ ions form ideal triangular lattice layers that are separated by the non-magnetic Na$Ch_6$ octahedra along the $c$ axis. It is notable that the Yb$Ch_6$ octahedra edge-shared with Na$Ch_6$ octahedra and are weakly distorted. The tilt of the YbSe$_6$ octahedra is smaller than that of YbS$_6$ octahedra, which is related to the larger radius of Se$^{2-}$ and leads to the difference of intralayer interactions between NaYbSe$_2$ and NaYbS$_2$. Due to the large difference in ionic size between Na$^+$ and Yb$^{3+}$, the anti-site disorder is much smaller than that of the Yb-based QSL candidate YbMgGaO$_4$ with Mg$^{2+}$/Ga$^{3+}$ site disorder \cite{NC4949}. The site disorder in NaYbS$_2$ and NaYbSe$_2$ has been well studied. For NaYbS$_2$, the results of X-ray diffraction (XRD), scanning electron microscopy (SEM), electron spin resonance (ESR), and nuclear magnetic resonance (NMR) all evidenced an absence of inherent structural distortions \cite{PRB220409, QuantumFrontiers13}. For NaYbSe$_2$, the previous literatures have reported the absence of intrinsic structural disorder by using XRD, ESR, NMR, and neutron scattering measurements \cite{PRB224417, Scheie}. An exceptional result is that a small amount of Yb occupying the Na site was probed by using single-crystal XRD and the inductively coupled plasma measurements \cite{PRX021044}. The structural and compositional characterization of these single crystals have been reported in some of our previous papers \cite{CPL35, PhysRevB184419, SciChinaPhysMechAstron67}. However, we have not carried out deeper experimental characterization of the site disorder in the samples used in the present work.

\begin{figure}
\includegraphics[clip,width=6cm]{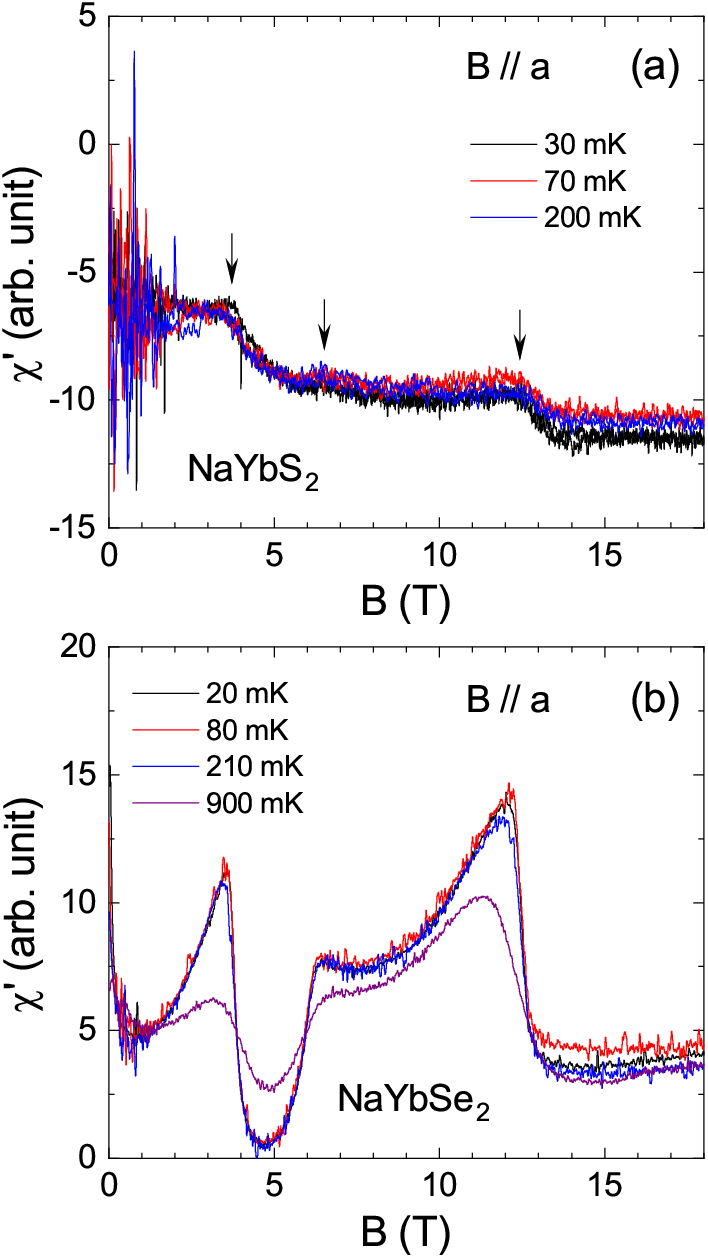}
\caption{Real component of the ac magnetic susceptibility as a function of dc field at different temperatures with $B \parallel a$ for NaYbS$_2$ and NaYbSe$_2$, respectively. The arrows indicate the weak peaks of NaYbS$_2$ data.}
\label{M}
\end{figure}

Figure 2(a) shows the field dependence of ac susceptibility $\chi'(B)$ measured at selected temperatures with $B \parallel a$ for NaYbS$_2$. At $T =$ 30 mK, the $\chi'(B)$ curve displays three weak peaks around 3.5, 6.5, and 12.5 T, respectively, and approaches saturation around 13.5 T. With increasing temperature, these peaks become weaker and broader. These transition fields are consistent with the previously reported results, indicating the magnetic transitions induced by the external magnetic field \cite{QuantumFrontiers13}. For NaYbSe$_2$, at $T =$ 20 mK, the $\chi'(B)$ curve exhibits much sharper peaks around 3.5, 6.5, and 12 T with $B \parallel a$. With increasing temperature, these peaks become broader and shift to lower magnetic field, as shown in Fig. 2(b). The overall behavior of $\chi'(B)$ curves are similar to those of NaYbS$_2$. The remarkable difference is that there is a deep valley at 3.5 $\sim$ 6.5 T for NaYbSe$_2$, which corresponds to the 1/3$M_s$ plateau ($M_s$ is the saturation magnetization), as reported by Ranjith {\it et al.} \cite{PRB224417}, and indicates an up-up-down phase. This has been predicted by the mean-field theory and further confirms an easy-plane anisotropy in NaYbSe$_2$ \cite{CondensMatter69}. Therefore, the present ac susceptibility data are basically consistent with the previous magnetization studies. In passing, we explain a bit about the difference of ac susceptibility data between NaYbS$_2$ and NaYbSe$_2$. The ac susceptibility measurement was performed using home-made ac coils \cite{PRB064401}. Different coils lead to different background of ac signal. The data of NaYbS$_2$ and NaYbSe$_2$ were obtained on two different coils and the coil for NaYbS$_2$ has large background. This is the main reason for that there is no so clear valley feature of NaYbSe$_2$ data.

While completing the present work, we noticed a recent study on NaYbSe$_2$ by Scheie {\it et al.} that reported the ultralow-temperature ac susceptibility for not only $B \parallel a$ but also $B \parallel c$ \cite{Scheie}. It can be seen that our susceptibility data are in good consistent with theirs. It should be pointed out that their data for $B \parallel c$ show a very broad and weak peak-like feature at $\sim$ 7 T, which behaved significantly different from those for $B \parallel a$.


\subsection{Specific heat}

\begin{figure}
\includegraphics[clip,width=6cm]{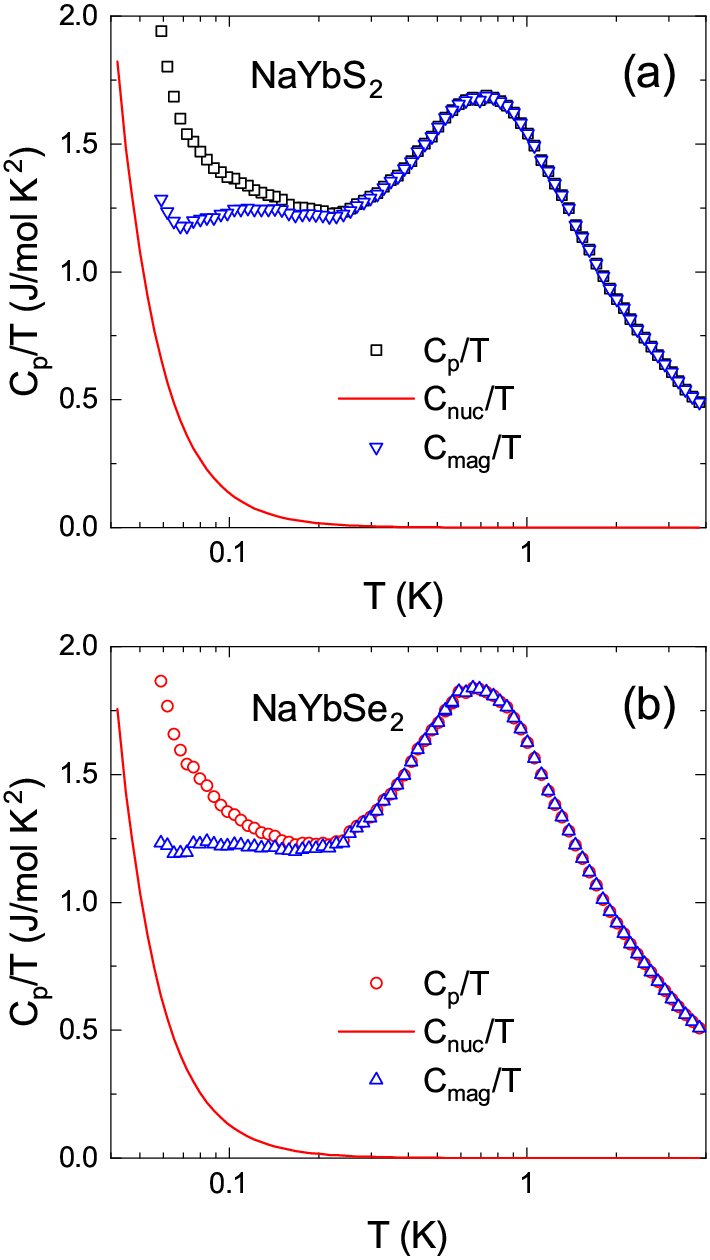}
\caption{Temperature dependence of the specific heat divided by temperature, $C_p/T$, for NaYbS$_2$ and NaYbSe$_2$ single crystals at zero magnetic field. The red lines are the estimated nuclear Schottky contributions $C_{\rm{nuc}}/T$. Also shown are the magnetic specific heat divided by temperatures, $C_{\rm{mag}}/T$.}
\label{Cp}
\end{figure}

To characterize the thermodynamics of NaYbS$_2$ and NaYbSe$_2$ single crystals, the low-temperature specific heat measurements were performed down to 60 mK with external magnetic field along the $c$ axis. As shown in Fig. 3, the zero-field specific heat is almost identical for NaYbS$_2$ and NaYbSe$_2$, and there is no signature of long-range magnetic order. The upturn at ultralow temperatures is originated from the nuclear Schottky contribution, which can be simply described by $C_{\rm{nuc}} \propto T^2$ \cite{PRB224417, PRX021044}. The plot $C_p/T$ vs $T$ displays a broad peak between 0.2 and 1.5 K, which is a common behavior observed in some other QSL candidates, such as YbMgGaO$_4$ and NaYbO$_2$ \cite{NatPhys15, SciRep5}. After subtracting the estimated nuclear Schottky contributions, the obtained magnetic specific heat $C_{\rm{mag}}$ show a nearly linear temperature dependence at low temperatures, which indicates a gapless QSL ground state with a constant density of states. All these results are reasonably consistent with the previous reports \cite{PhysRevB184419, QuantumFrontiers13, PRX021044}. Moreover, the constant density of states at low energies can be explained by the spinon Fermi surface in QSL. Although the linear-$T$ specific heat and the constant density of states can be understood in terms of the random field effect in the spin glass, we think this is unlikely for our systems here. There does not exist any signature of spin freezing transition in our measurement, and thus the spin glassy features are absent. The Yb$^{3+}$ ions in these systems provide the Kramers doublet whose degeneracy is protected by the time reversal symmetry. Unlike the non-Kramers doublets for which the random fields can be obtained by the crystal disorders, the non-magnetic disorders cannot generate the random fields for the Kramers doublets. Thus, we think the spinon Fermi surface state is more compatible with the low-temperature specific heat behaviors.

\begin{figure}
\includegraphics[clip,width=6cm]{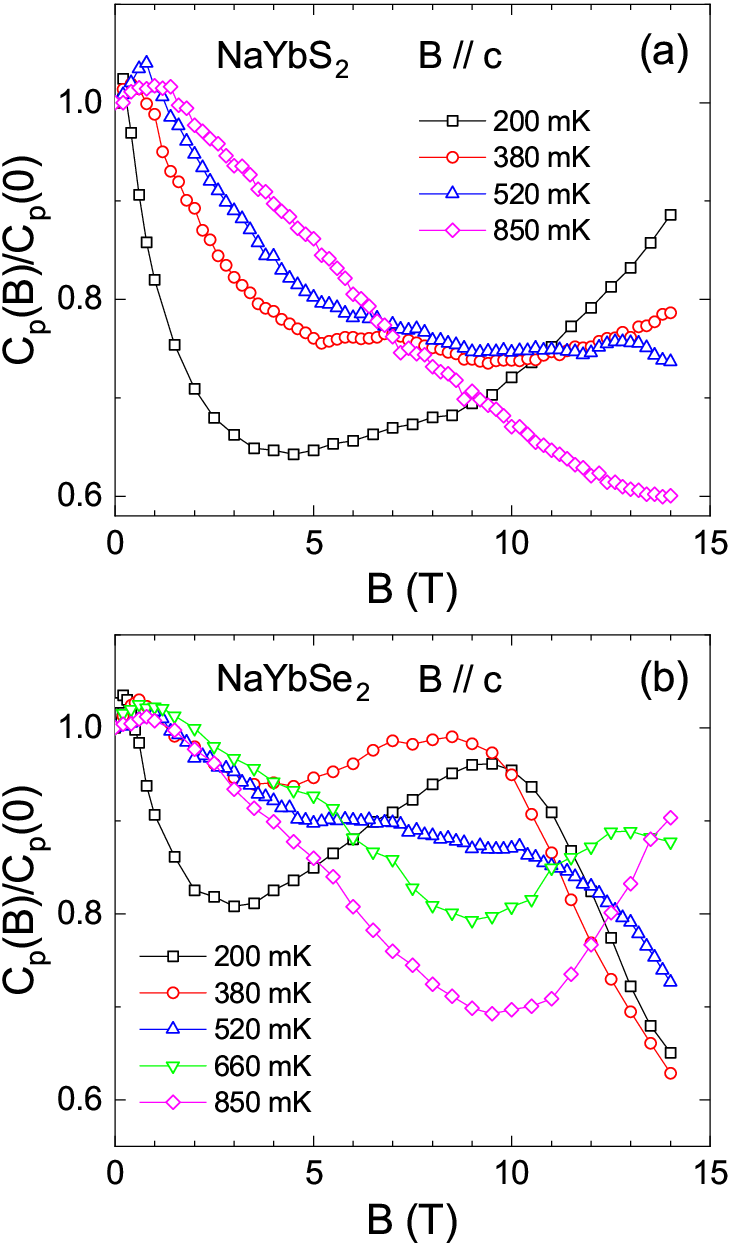}
\caption{Magnetic field dependence of specific heat for NaYbS$_2$ and NaYbSe$_2$ single crystals with $B \parallel c$.}
\label{Cp-H}
\end{figure}

Figures 4(a) and 4(b) show the magnetic field dependencies of specific heat of NaYbS$_2$ and NaYbSe$_2$ at selected temperatures with external magnetic field along the $c$ axis, respectively. There are some similarities between the data of these two materials: (i) there is a weak peak at low fields, which shifts to higher magnetic field with increasing temperature and is related to the nuclear Schottky anomaly. (ii) With increasing magnetic field, some field-induced transitions appear while the overall trend is decreasing. There are also certain differences between these two materials. At $T =$ 200 mK, the $C_p(B)/C_p(0)$ of NaYbS$_2$ first decreases and then increases with increasing magnetic field, exhibiting a broad valley and a kink around 4 and 10 T, respectively. However, at $T =$ 200 mK, the $C_p(B)/C_p(0)$ curve of NaYbSe$_2$ exhibits a valley and a broad peak around 3 and 9 T, respectively; and the specific heat continues to decrease at high field, which is contrast to the case of NaYbS$_2$. In addition, for NaYbSe$_2$, the low-field valley becomes weaker and shifts to higher magnetic field with increasing temperature. It is likely that these intermediate-field behaviors are correlated with the broad and weak peak-like feature of ac susceptibility data \cite{Scheie}. It should be pointed out that the nuclear Schottky term can also be strongly affected by the magnetic field; in particular, at very low temperatures it can be strongly suppressed upon increasing field.

\subsection{Thermal conductivity}

\begin{figure}
\includegraphics[clip,width=6.5cm]{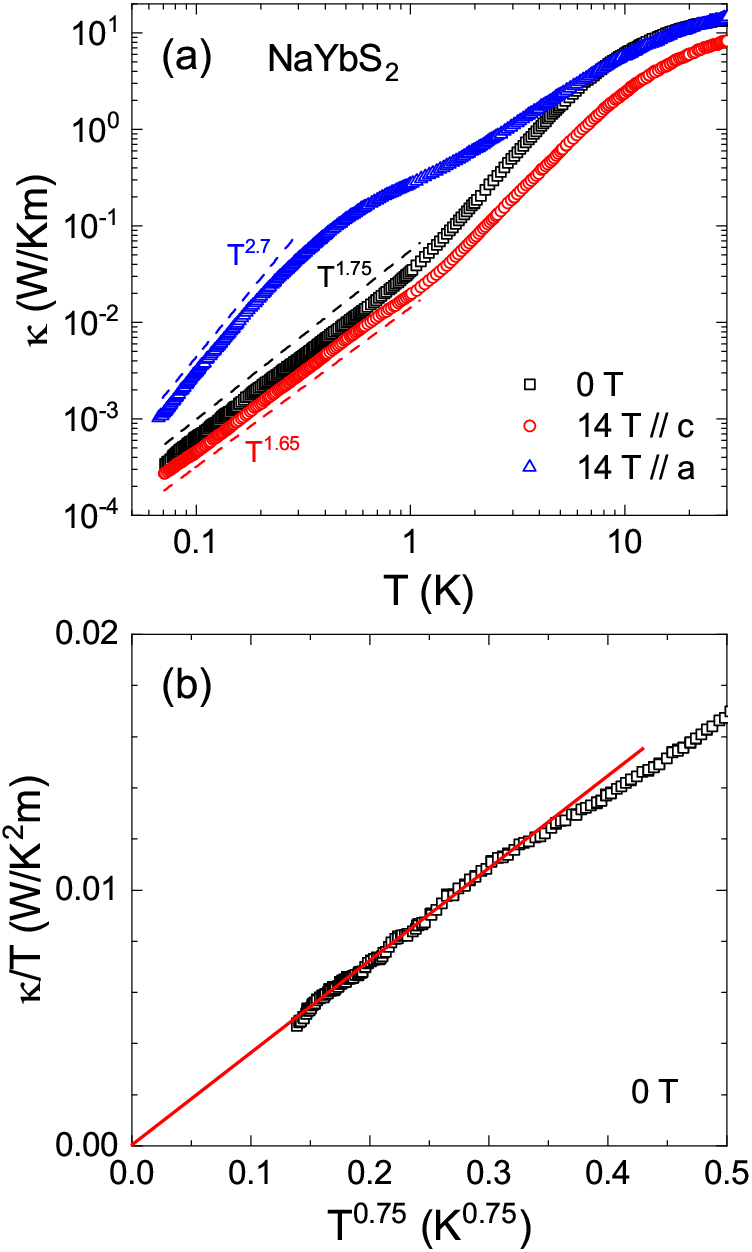}
\caption{(a) Temperature dependence of the thermal conductivity of NaYbS$_2$ single crystal in zero and 14 T magnetic field along the $c$ axis or the $a$ axis. The dash lines indicate the power law temperature dependence. (b) Zero-field thermal conductivity data plotted as $\kappa/T$ vs $T^{0.75}$ at very low temperatures and the solid line shows the linear fitting. There is a negligibly small residual term ($\kappa_0/T =$ 0).}
\label{kT}
\end{figure}

Figure 5(a) shows the temperature dependence of thermal conductivity for NaYbS$_2$ single crystal in zero field and in 14 T magnetic field along the $c$ axis or the $a$ axis. As one can see, the zero-field $\kappa(T)$ curve exhibits a $T^{1.75}$ power law behavior at very low temperatures, which distinctly deviates from the standard phonon transport behavior ($T^3$-dependence) \cite{Berman}. This deviation is apparently due to the strong phonon scattering by magnetic excitations. In addition, the direction of the external magnetic field has a significantly different effect on changing thermal conductivity. To be more specific, the 14 T field along the $c$ axis slightly suppresses the thermal conductivity and the $\kappa(T)$ curve roughly obeys a $T^{1.65}$ dependence. On the contrary, the thermal conductivity is strongly enhanced for 14 T $\parallel a$ and displays a rough $T^{2.7}$ dependence at low temperatures, which is quite close to the typical behavior of phonon thermal conductivity in the boundary scattering limit. The previous studies on the magnetization of NaYbS$_2$ single crystal revealed that 14 T along the $a$ axis can nearly fully polarize the Yb$^{3+}$ spins, while the spin polarization field is much higher for the $c$ direction \cite{QuantumFrontiers13, PRB220409}. Therefore, under the 14 T $c$-axis field the spins are still in a fluctuating state and can strongly scatter phonons, while the 14 T in-plane field can almost smear out the magnetic scattering on the phonons. Figure 5(b) shows the ultralow-temperature thermal conductivity at zero field. We fit the data by using a formula $\kappa/T = $ $\kappa_0/T$ + $bT^{\alpha-1}$ \cite{Science328, NC4216, NC4949}, in which the two terms represent contributions from the itinerant fermionic excitations and phonons, respectively. The fitting gives $\kappa_0/T$ = 0 and $\alpha =$ 1.75. The zero residual term implies the absence of the itinerant fermionic excitations in NaYbS$_2$.

\begin{figure}
\includegraphics[clip,width=8.5cm]{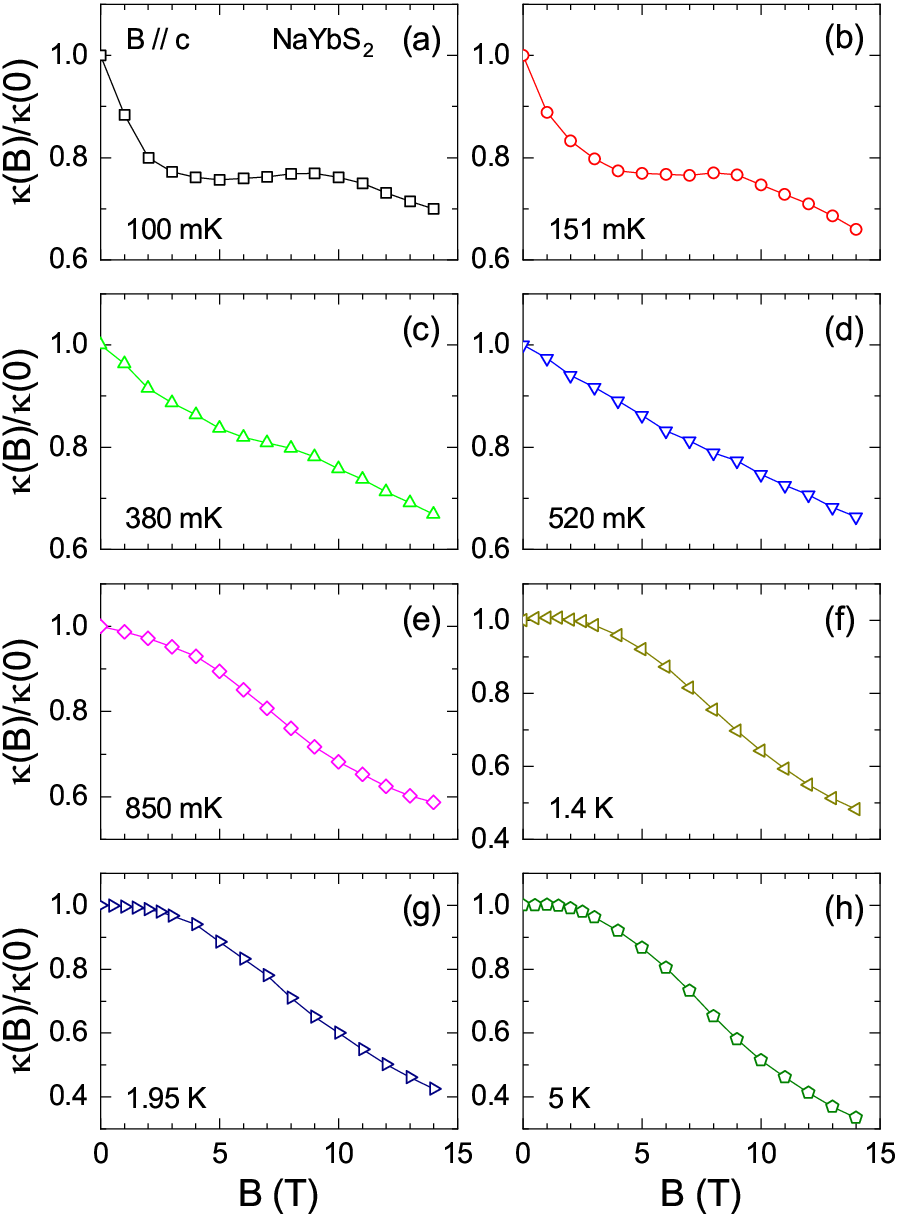}
\caption{Magnetic-field dependencies of thermal conductivity of NaYbS$_2$ single crystal at select temperatures for $B \parallel c$.}
\label{kH}
\end{figure}

\begin{figure}
\includegraphics[clip,width=8.5cm]{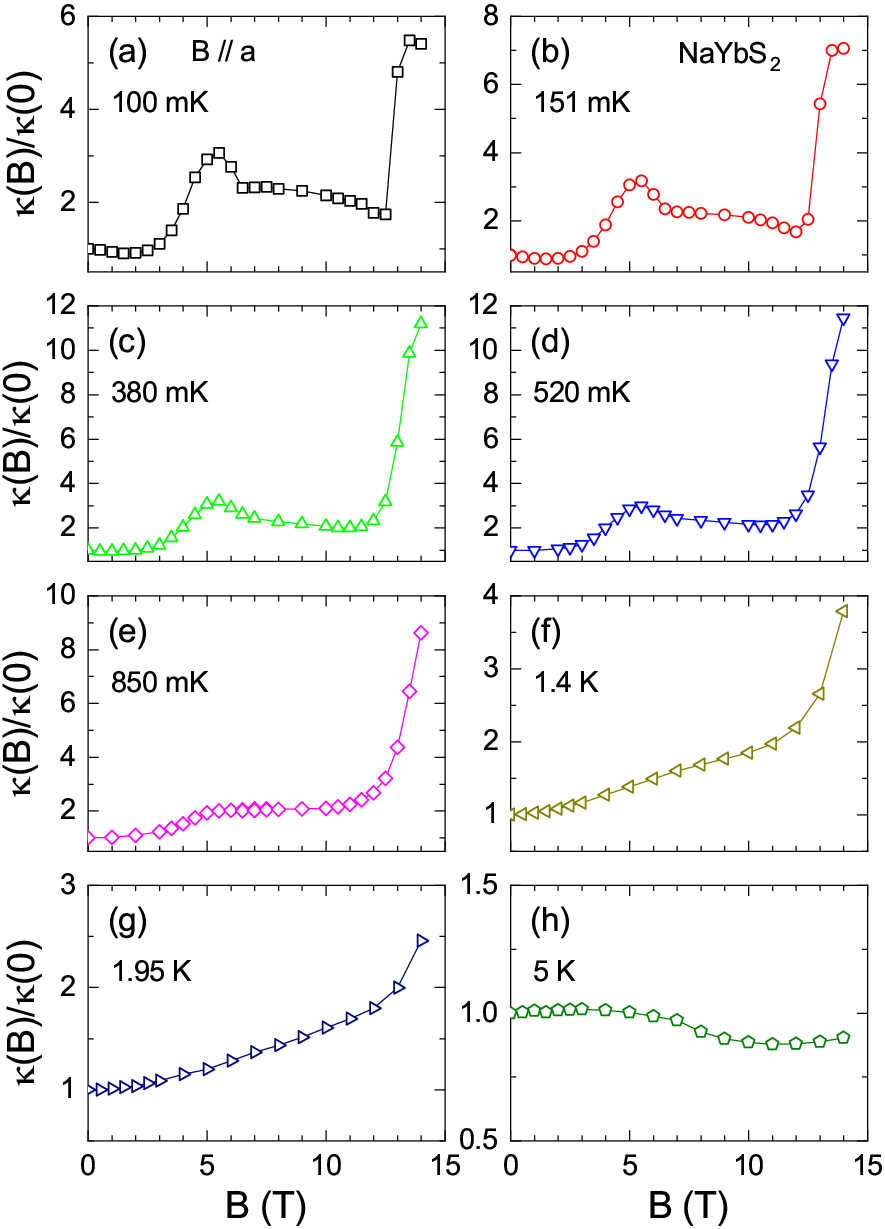}
\caption{Magnetic-field dependencies of thermal conductivity of NaYbS$_2$ single crystal at select temperatures for $B \parallel a$.}
\label{kH}
\end{figure}

Figures 6 and 7 show the $\kappa(B)$ isotherms for NaYbS$_2$ single crystal with $B \parallel c$ and $B \parallel a$, respectively. It is notable that the field dependence of $\kappa$ are significantly different between the $c$-axis and in-plane fields. For $B \parallel c$, the $\kappa$ is always suppressed by magnetic field and the suppression can reach $\sim$ 40$\%$ at 14 T without saturation. At very low temperatures of 100 and 151 mK, the $\kappa(B)/\kappa(0)$ curves show a shallow and broad dip at $\sim$ 5 T and an anomaly around 9 T, which disappear with increasing temperature and is consistent with the anomalies in the specific heat. In contrast, the in-plane field induces much more complex $\kappa(B)$ behavior with strong enhancement of $\kappa$ at higher fields. These indicate the strong anisotropic behavior between the $c$-axis and in-plane fields. At very low temperatures, upon increasing field the $\kappa$ firstly decreases slightly ($\sim$ 10 \%) and then increases non-monotonically. The $\kappa(B)/\kappa(0)$ curves display a shallow valley around 2 T, a broad peak between 3 and 6 T, and a sharp increase around 12.5 T, accompanied with a saturation above 13.5 T. At higher temperatures, the low-field broad peak weakens and the 14 T-field enhancement can reach $\sim$ 10 times without saturation. The previous magnetization and torque measurements indicated that there are four transition fields around 3.3, 6.1, 10.2, and 14.8 T for $B \parallel a$ at $T =$ 0.8 K \cite{QuantumFrontiers13}. Therefore, the characteristic fields at $\kappa(B)/\kappa(0)$ isotherms are roughly consistent with these transition fields by magnetic measurements. The broad peak is related to the up-up-down phase and nearly disappears at $T =$ 850 mK. The sharp increase of $\kappa$ is associated with a phase transition from the up-up-down phase to the oblique phase. Moreover, the saturation magnetic field is reached around 13.5 T at low temperatures. At higher temperatures above 151 mK, the $\kappa(B)/\kappa(0)$ curves may saturate at fields higher than 14 T, which is consistent with higher saturation field of magnetization \cite{QuantumFrontiers13}.

\begin{figure}
\includegraphics[clip,width=6.5cm]{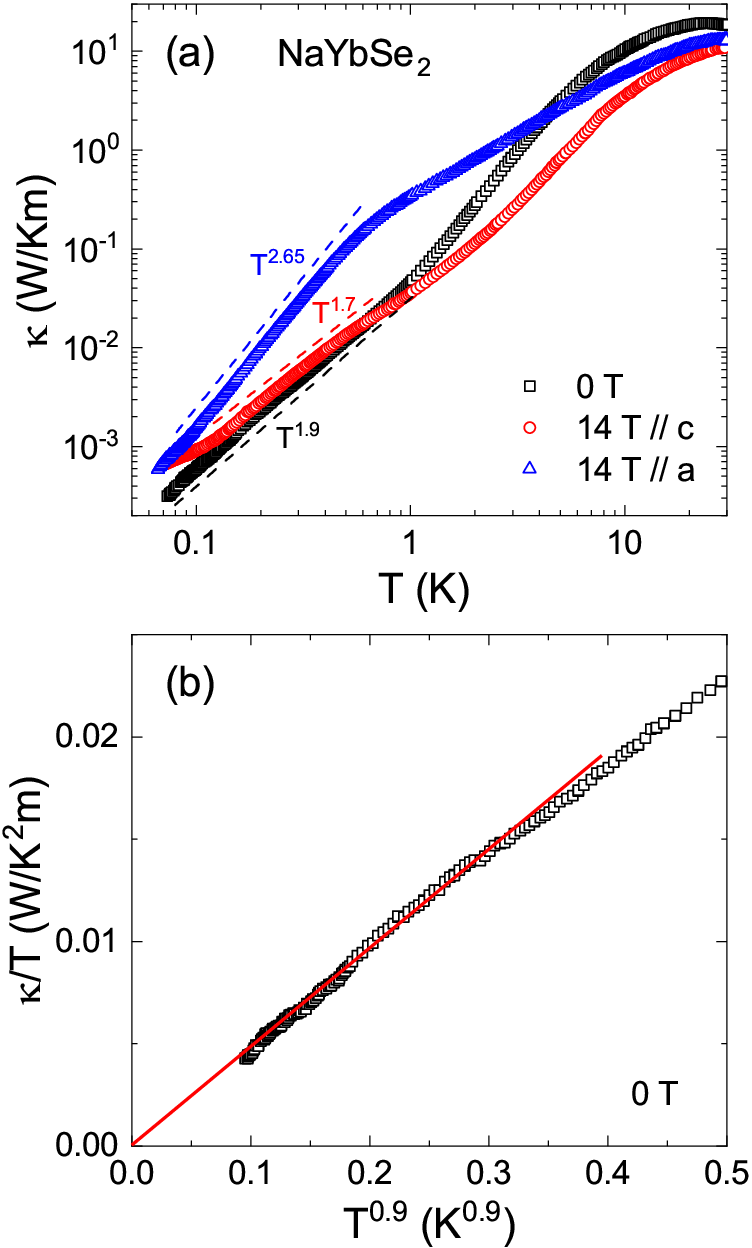}
\caption{(a) Temperature dependence of the thermal conductivity of NaYbSe$_2$ single crystal in zero and 14 T magnetic field along the $c$ axis or the $a$ axis, respectively. The dash lines indicate the power law temperature dependence. (b) Zero-field thermal conductivity plotted as $\kappa/T$ vs $T^{0.9}$ at very low temperatures and the solid line shows the linear fitting. There is a negligibly small residual term ($\kappa_0/T =$ 0).}
\label{kappaT}
\end{figure}

Figure 8(a) shows the temperature dependence of thermal conductivity for NaYbSe$_2$ single crystal in zero field and in 14 T magnetic field along the $c$ axis or the $a$ axis. In zero field, the $\kappa(T)$ curve exhibits a rough $T^{1.9}$ dependence, which also deviates from the standard $T^3$ behavior of phonon thermal conductivity at the boundary scattering limit \cite{Berman} and indicates strong phonon scattering. Moreover, we observed a negligibly small $\kappa_0/T$ in the $\kappa/T$ vs $T^{0.9}$ plot, as shown in Fig. 8(b), which also points to the absence of mobile fermionic excitations. The different responses are observed for applying 14 T field along the $c$ or $a$ axis. The $\kappa(T)$ curve displays a $T^{2.65}$ dependence for 14 T field along the $a$ axis, which is slightly weaker than the standard $T^3$ behavior and may be due to the specular reflections at the sample surfaces or the remaining spin-phonon scattering. For 14 T field along the $c$ axis, the $\kappa(T)$ curve displays shows a $T^{1.7}$ dependence and even weaker temperature dependence below 100 mK, which also indicates the strong phonons scattering by the magnetic excitations.

\begin{figure}
\includegraphics[clip,width=8.5cm]{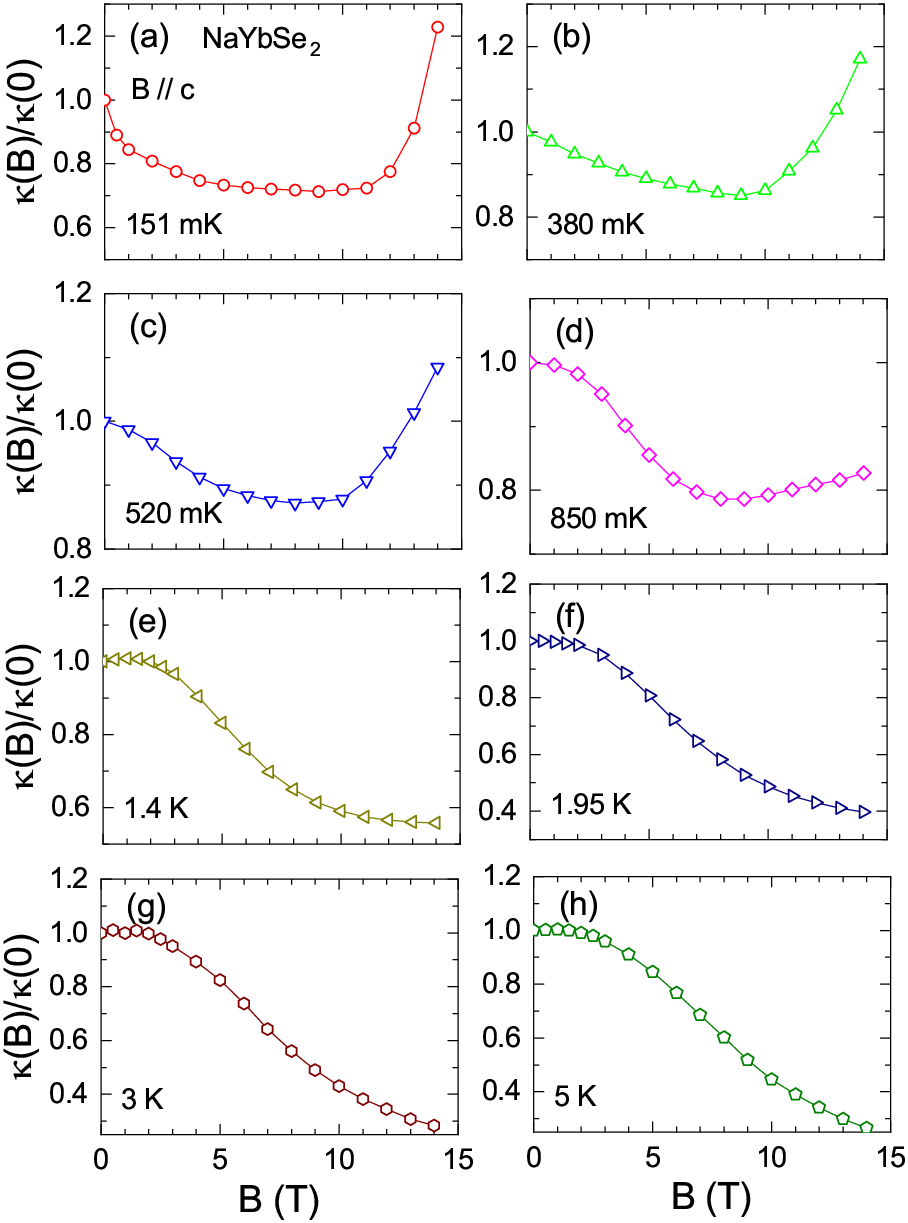}
\caption{Magnetic-field dependencies of thermal conductivity of NaYbSe$_2$ single crystal at select temperatures for $B \parallel c$.}
\label{kappaH}
\end{figure}

\begin{figure}
\includegraphics[clip,width=8.5cm]{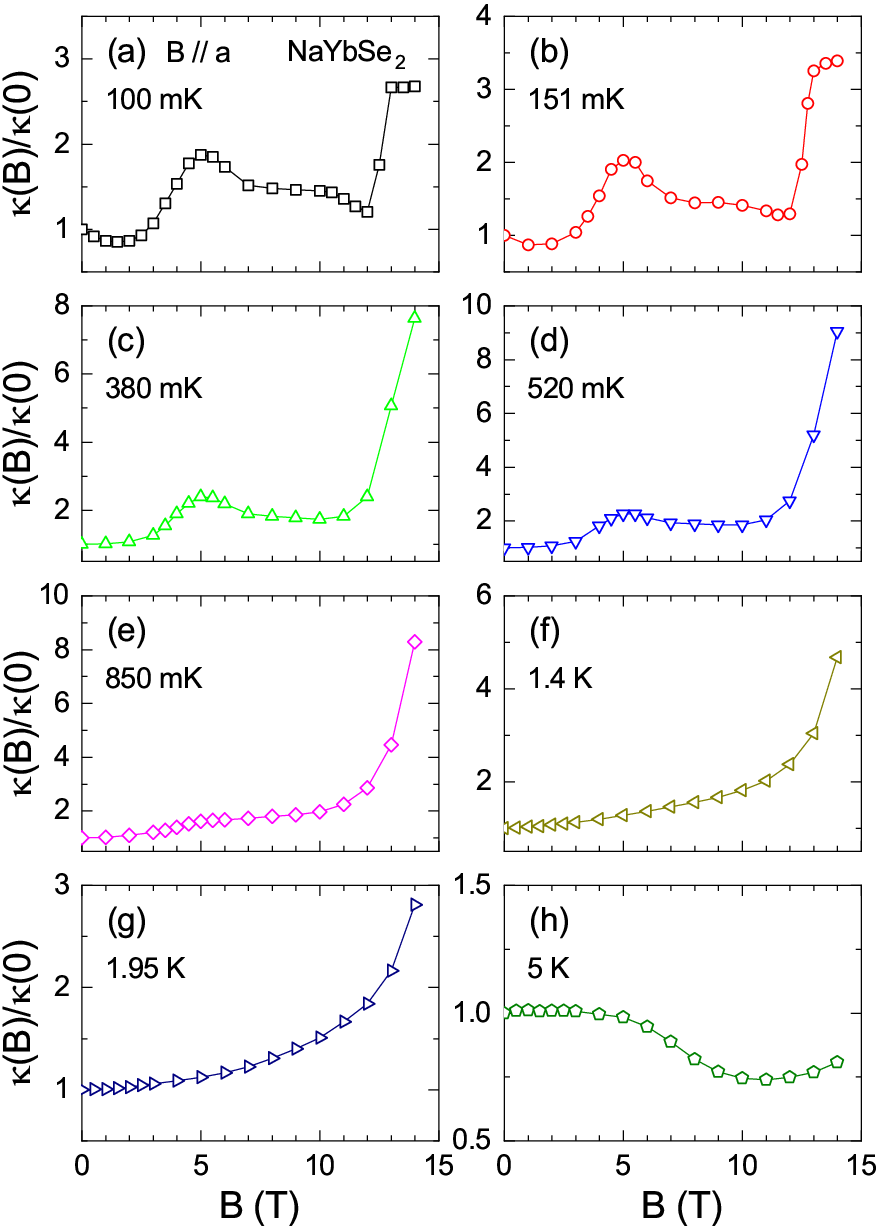}
\caption{Magnetic-field dependencies of thermal conductivity of NaYbSe$_2$ single crystal at select temperatures for $B \parallel a$.}
\label{kappaH}
\end{figure}

The $\kappa(B)/\kappa(0)$ isotherms of NaYbSe$_2$ single crystal for $B \parallel c$ and $B \parallel a$ are shown in Figs. 9 and 10, respectively. It can be seen that the $\kappa(B)/\kappa(0)$ curves exhibit a broad valley at low fields followed by an enhancement at high fields for $B \parallel c$. The $\kappa(B)/\kappa(0)$ for $B \parallel a$ behaves very similar to that of NaYbS$_2$, and there is a broad peak around 4 $\sim$ 6 T, corresponding to the up-up-down spin arrangement, which is consistent with 1/3$M_s$ plateau. Moreover, a shallow dip can be observed around 12 T, which is a field-induced magnetic transition, and the $\kappa(B)$ curves finally saturate at 13 T for $T =$ 100 and 151 mK. With increasing temperature, the broad peak becomes invisible and the shallow dip gradually disappears. All these critical fields are consistent with the previous magnetization results \cite{PRB224417}.

\section{Discussion}

It was reported that NaYbS$_2$ and NaYbSe$_2$ evidenced of gapless spin excitations (spinons) from specific heat, inelastic neutron scattering (INS) and muon spin relaxation ($\mu$SR) measurements, which are connected to QSL with a spinon Fermi surface \cite{QuantumFrontiers13, PRX021044, PRB085115}. The itinerant gapless spinons are expected to contribute to the thermal conductivity, which yields a nonzero residual linear term $\kappa_0/T$ at $T \to$ 0 \cite{NC4949, Science328, NC4216}. It has become the smoking gun of the itinerant spinons and gapless QSL. However, the negligibly small $\kappa_0/T$ was observed at zero-Kelvin limit in present work. Thus, the spinons may be strongly scattered by phonons or a minority of quasi-static spins in a fluctuating state, which was detected by $\mu$SR and NMR \cite{TheInnovation, PRB241116R}. Based on the magnetic field dependence of $\kappa$, it is more likely that there are low-energy spinons scattering with phonons. It is worthy of pointing out that in most of QSL candidates, there are rather strong coupling between phonons and spinons, which leads to weak temperature dependence of ultralow-temperature $\kappa$ and rather small $\kappa_0/T$ contributed by the spinon transport. Therefore, even though the present experiments display zero $\kappa_0/T$ at zero field, the $\kappa$ data actually indicate the existence of spinons that scatter with phonons. This possibility was recently found also in Ising-type QSL candidate PrMgAl$_{11}$O$_{19}$ \cite{PrMgAl11O19}. It is worthy of pointing out that the previous ultralow-temperature $\kappa$ measurement of NaYbSe$_2$ single crystal also show the absence of $\kappa_0/T$ \cite{TheInnovation}, although the magnetic field dependence of $\kappa$ is very different from ours. Very recently, inelastic neutron scattering experiments and ac susceptibility down to 20 mK \cite{Scheie} suggest that there exists a very tiny gap in NaYbSe2. This may offer an alternative possibility to understand our seemingly paradoxical data.

As one can see, the overall behavior of $\chi'(B)$ curves and $\kappa(B)/\kappa(0)$ curves for $B \parallel a$ are highly similar between NaYbS$_2$ and NaYbSe$_2$, which indicates the similar in-plane properties of them with easy-plane anisotropy. However, there is a significant difference in the $\kappa(B)/\kappa(0)$ behavior with $B \parallel c$ between NaYbS$_2$ and NaYbSe$_2$. For NaYbS$_2$, the low-temperature $\kappa(B)/\kappa(0)$ curves continue to decrease as the magnetic field increases, while the low-temperature $\kappa(B)/\kappa(0)$ curves of NaYbSe$_2$ show a reduction at low field followed by an enhancement at high fields. It is seems that in both materials the spin excitations (likely spinons) strongly scatter phonons rather than transport heat. High enough magnetic field along either the $c$ axis or the $a$ axis would suppress the spin excitations and significantly recover the phonon thermal conductivity. For $B \parallel c$, the continuous decrease of $\kappa$ with field in NaYbS$_2$, which is different from the high-field increase of $\kappa$ in NaYbSe$_2$, is likely due to that 14 T is still far from the polarization field in NaYbS$_2$. Nevertheless, the low-field anomaly $\kappa(B)/\kappa(0)$ with $B \parallel c$ in NaYbS$_2$ may suggest some unknown field-induced transition, which is absent in NaYbSe$_2$.

\section{Summary}

In summary, we have successfully grown high-quality single crystals of the Yb-based triangular lattice QSL candidates NaYbS$_2$ and NaYbSe$_2$, and perform the ultralow temperature ac susceptibility, specific heat, and thermal conductivity measurements with external magnetic field along the $c$ axis or the $a$ axis. For $B \parallel a$, the $\chi'(B)$ and $\kappa(B)$ isotherms display similar field-induced magnetic transitions for these two materials, while the $C_p(B)$ and $\kappa(B)$ isotherms exhibit significant difference between NaYbS$_2$ and NaYbSe$_2$ for $B \parallel c$. The magnetic specific heat exhibit a nearly linear temperature dependence at $T <$ 200 mK. However, the ultralow temperature thermal conductivity shows a negligibly residual term $\kappa_0/T$. The temperature and field dependence of $\kappa$ further indicate that there are spinon excitations scattering phonons rather than transporting heat in NaYbS$_2$ and NaYbSe$_2$. These results point to the gapless or tiny-gapped QSL ground state of these materials.

\begin{acknowledgements}

This work was supported by the National Key Research and Development Program of China (Grant No. 2023YFA1406500 and 2022YFA1402704), the National Natural Science Foundation of China (Grant Nos. 12404043, 12274388, 12274186, 12174361, 12104010, and 12104011), the Nature Science Foundation of Anhui Province (Grant Nos. 1908085MA09 and 2108085QA22), the Strategic Priority Research Program of the Chinese Academy of Sciences (Grant No. XDB33010100), and the Synergetic Extreme Condition User Facility (SECUF). The work at the University of Tennessee was supported by the National Science Foundation through award DMR-2003117. A portion of this work was performed at the National High Magnetic Field Laboratory, which is supported by the National Science Foundation Cooperative Agreement No. DMR-1644779 and the State of Florida.

\end{acknowledgements}

\end{document}